\documentclass[aps,prl,reprint,groupedaddress]{revtex4-2}

\usepackage{natbib}
\usepackage[percent]{overpic}
\usepackage{xcolor}
\usepackage{amsmath}
\usepackage{hyperref}
\newcommand\solphys{Sol.~Phys.}%
\newcommand\aap{Astron.~Astrophys.}%
\newcommand\apjl{ Astrophys.~J.~Lett.}%
\newcommand\mnras{MNRAS}%

\graphicspath{{./}{./figs/}}
\begin{document}
\title{Evidence for Vortex Shedding in the Sun's Hot Corona}
\author{Tanmoy Samanta$^{1}$}
\author{Hui Tian$^{1}$}
\email[]{huitian@pku.edu.cn}
\author{Valery M. Nakariakov$^{2,3}$}
\affiliation{$^{1}$ School of Earth and Space Sciences, Peking University, Beijing 100871, China}
\affiliation{$^{2}$ Centre for Fusion, Space and Astrophysics, University of Warwick, Coventry, CV47AL, United Kingdom}
\affiliation{$^{3}$ St. Petersburg Branch, Special Astrophysical Observatory, Russian Academy of Sciences, 196140, St. Petersburg, Russia}

\date{\today}

\begin{abstract}
Vortex shedding is an oscillating flow that is commonly observed in fluids due to the presence of a blunt body in a flowing medium. Numerical simulations have shown that the phenomenon of vortex shedding could also develop in the magnetohydrodynamic (MHD) domain. The dimensionless Strouhal number, the ratio of the blunt body diameter to the product of the period of vortex shedding and the speed of a flowing medium, is a robust indicator for vortex shedding, and, generally of the order of 0.2 for a wide range of Reynolds number. Using an observation from the Atmospheric Imaging Assembly  on board the Solar Dynamics Observatory, we report a wavelike or oscillating plasma flow propagating upward against the Sun's gravitational force. A newly formed shrinking loop in the post-flare region possibly generates the oscillation of the upflow in the wake of the hot and dense loop through vortex shedding. The computed Strouhal number is consistent with the prediction from previous MHD simulations. Our observation suggests the possibility of vortex shedding in the solar corona. 
\end{abstract}

\pacs{{96.60.Hv}, {96.60.Iv}, {96.60.ph}, {96.60.qd}, {96.60.qe}, {96.60.qf}}

\maketitle

It is well known that the interaction of a steady flow with an obstacle in fluids could generate a sequence of vortices just behind the obstacle. This hydrodynamic phenomenon is known as the von K$\acute{\text{a}}$rm$\acute{\text{a}}$n vortex street or the vortex shedding \citep{Tritton1977,Williamson1996}. The vortices are produced periodically with opposite vorticity from the two sides of the obstacle and are dragged by the flow. The phenomenon of vortex shedding has been widely studied in both  science and engineering in hydrodynamic conditions. For example, the waving of a flag in the wind is due to the vortex shedding effect. It is also reported that tall chimneys in the presence of air flow could generate vortex shedding, which can lead to violent oscillation and damaging of the chimney. 
The behavior of a fluid in the presence of a blunt obstacle can be described by the Reynolds number, $R = VL$/${\nu}$, and the Strouhal number, $St = L$/$(PV)$, where V, L, $\nu$, and P are the flow speed, size of the blunt body, kinematic viscosity, and period of the vortex shedding, respectively.

The outer atmosphere of our Sun, the million-degree corona, is highly structured and strongly coupled to the magnetic field. A wide range of dynamical phenomena have been observed in the magnetohydrodynamic (MHD) environment of the corona. One numerical simulation  \citep{2010PhRvL.105e5004G} examined the phenomenon of vortex shedding in MHD conditions (called Alfv$\acute{\text{e}}$nic vortex shedding) and showed that periodic shedding of vortices occurs in a fashion similar to hydrodynamic or aerodynamic conditions. It has been suggested that this process may explain the excitation of some oscillations in coronal loops \citep{2009A&A...502..661N} and coronal mass ejections \citep{2015ApJ...803L...7L}. However, direct observational evidence of vortex shedding in a MHD environment is still missing \citep{2018A&A...615A.143N}.

\begin{figure*}
\label{fig1}
\begin{overpic}[width=\textwidth]{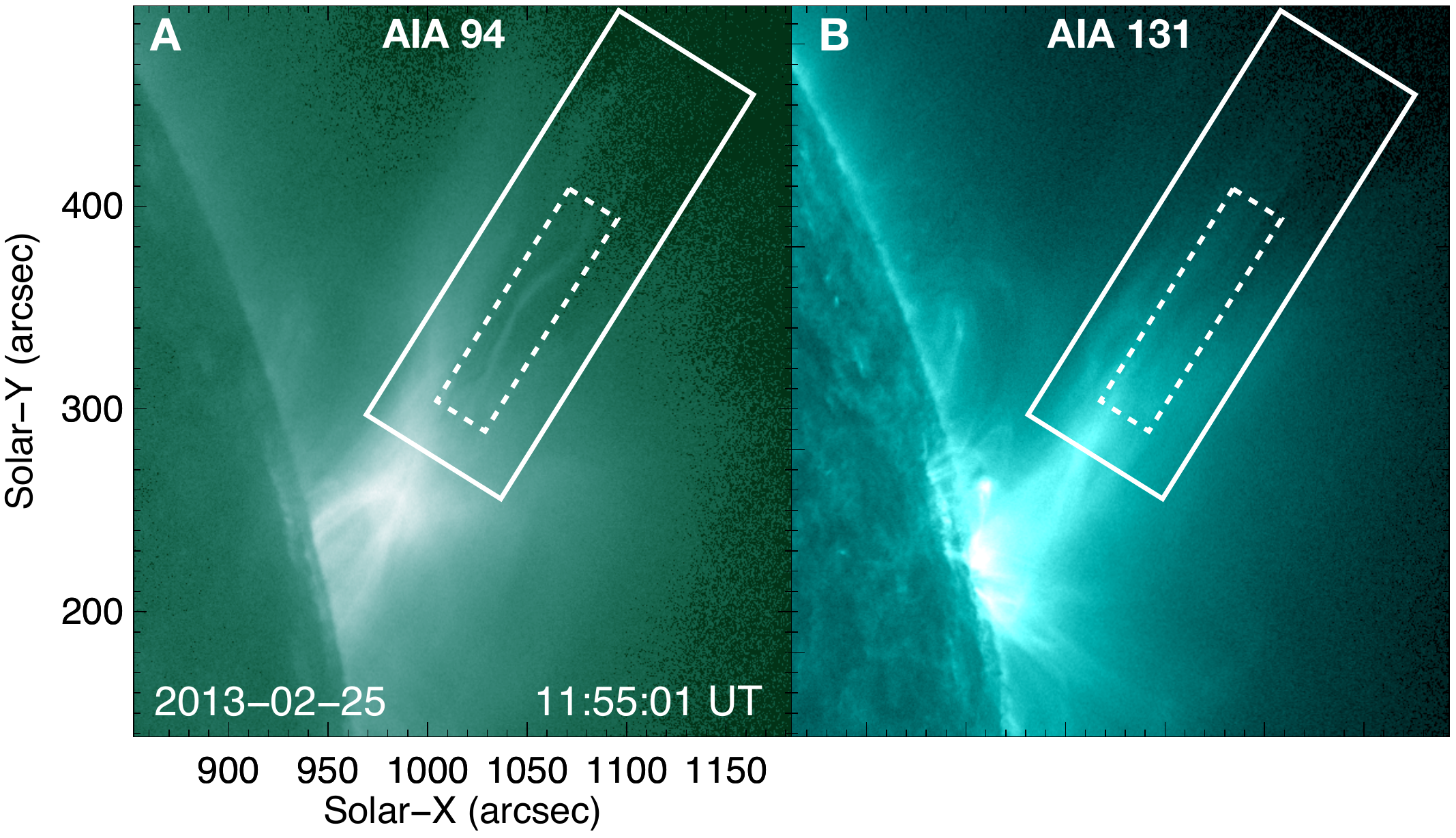}
\end{overpic}
\caption{The supra-arcade region observed in the SDO/AIA 94~\AA~and 131~\AA~filters at 11:55 UT. The big and small rectangular boxes indicate the field of view in Figs. 2A, 3A and 3B, respectively. An animation is associated with this figure.}
\end{figure*}

Here, we present possible observational evidence of vortex shedding in the magnetized solar atmosphere. The event was found in the so-called supra-arcade region of a solar flare. Solar flares are one of the most powerful energy release events in the Solar System, usually powered by magnetic reconnection in the solar corona. Supra-arcade regions refer to the hot diffuse regions  around the current sheets of magnetic reconnection, consisting of plasma with temperatures from a few million to a few tens of million kelvin \citep{1998SoPh..182..179S,1999ApJ...519L..93M,2000SoPh..195..381M,2003SoPh..217..247I,2011LRSP....8....6S,2014ApJ...786...95H,2015Sci...350.1238C,2017A&A...606A..84C,2018ApJ...854..122W}. They generally extend above the newly formed flare arcades \citep{2004ApJ...616.1224S} and can last for several hours. Often dark, tadpolelike structures called supra-arcade downflows (SADs) are seen to descend through the bright fans of the supra-arcade regions, leading to fingerlike shapes in the upper part of supra-arcade regions \citep{2000SoPh..195..381M, 2003SoPh..217..247I,2004ApJ...605L..77A,2009ApJ...697.1569M,2009MNRAS.400L..85C,2011ApJ...730...98S,2013ApJ...775L..14C,2014ApJ...796L..29G,2014ApJ...796...27I,2014ApJ...785..106K,2015ApJ...807....6C}. The SADs are possibly regions of low density and  high temperature \citep{2003SoPh..217..247I,2012ApJ...747L..40S,2017ApJ...836...55R} behind contracting flare loops (due to magnetic tension) \citep{2008ApJ...675..868R,2010ApJ...722..329S,2011ApJ...742...92W,2010ApJ...714L..41L,2011ApJ...730...98S}. They are best seen when the reconnection current sheets are face-on at the solar limb.

The event was observed during a B8.9 flare (classification based on soft X-ray flux measurements with the Geostationary Operational Environmental Satellite) occurring in an active region at the west limb of the Sun on February 25, 2013. The Atmospheric Imaging Assembly (AIA) \citep{2012SoPh..275...17L} instrument on board the Solar Dynamics Observatory (SDO) \citep{2012SoPh..275....3P}  imaged the evolution of this flare with different filters. All the AIA images were processed, coaligned and normalized with the standard \textit{aia\_prep.pro} routine available in SolarSoft. The AIA images were taken at a 12 s cadence in each extreme ultraviolet (EUV) filter, with a pixel size of $0.6''$. We averaged 5 image frames in each filter to increase the signal-to-noise ratio. Figure 1 shows the post-flare region imaged with the 94~\AA\ and 131~\AA\ filters, which are dominated by emission lines from the Fe XVIII and Fe XXI ions formed around 6 million kelvin and 10 million kelvin, respectively. Several dark voidlike structures (SADs) are prominently observed in the 131~\AA\ images. The SADs propagate apparently downward towards the flare arcades. Interestingly, an upward propagating structure with enhanced emission is clearly seen in the 94~\AA\ images. This upflow is absent in the other AIA passbands, indicating that the upflowing plasma has a temperature of $\sim$6 million kelvin. While propagating upward, this hot plasma flow reveals a clear wavelike shape. 

\begin{figure*}
\label{fig2}
\begin{overpic}[width=\textwidth]{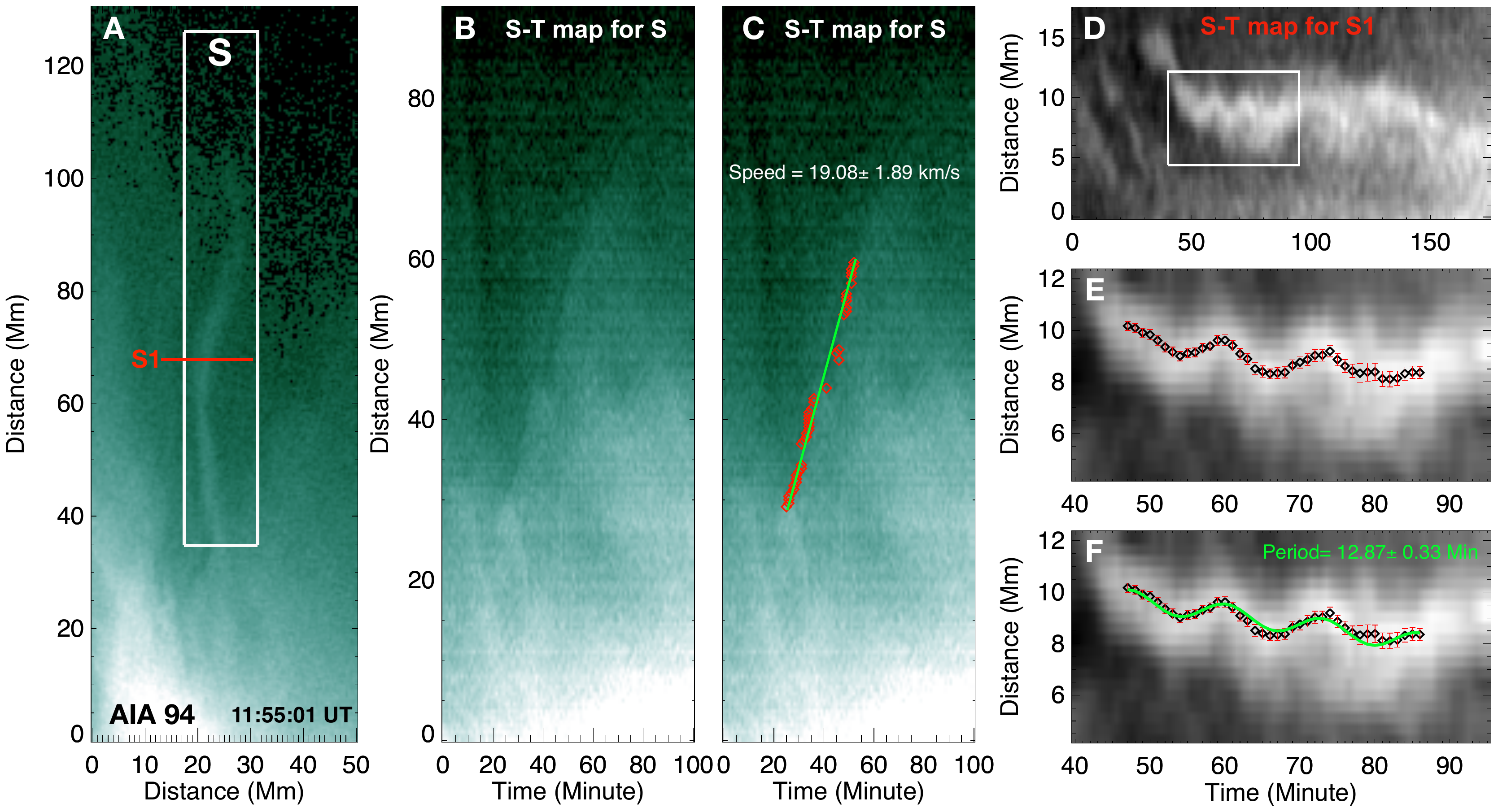}
\end{overpic}
\caption{Measurement of the physical parameters of the wavelike upflow. A: The upward propagating flow observed in the SDO/AIA 94~\AA\ channel at 11:55 UT. B: The space-time (S-T) diagram for the wide slit marked as S (rectangular box) in panel A. C: the background image is the same as panel B. The red diamonds mark the edge/ridge of the upflow in the S-T diagram. The upflow speed was estimated from the slope of the slanted straight line, which represents a linear fit to the red diamonds. D: The S-T diagram for the cut S1 shown in panel A. E-F: zoomed-in view of the white box in panel D. The black diamonds and red bars represent the central positions of the upflowing feature and the associated 1-$\sigma$ errors at different times, respectively. The period of the associated oscillation was determined through fitting with a sinusoidal function (the green line in panel F). An animation is associated with this figure.}
\end{figure*}

We have performed a quantitative analysis of the oscillating upflow and obtained several physical parameters of the flow. A wide slit, marked as S in Figure 2A, was chosen to cover the upflow during its entire lifetime. We averaged the AIA 94~\AA\ emission at each position along the slit and obtained a space-time (S-T) diagram, which is presented in Figure 2B. The S-T diagram clearly shows the upward propagation of the flow with time. The propagation speed can be estimated from the slope of the faint slanted ridge in the S-T diagram. First, we enhanced the edge of the slanted ridge using the \textit{sobel.pro} routine of \textit{IDL}. Afterwards, we determined the locations of the enhanced edge by finding the local maximum at each time step. We removed very faint edge detections by applying a threshold. The determined edge locations are marked by the red diamonds in Figure 2C. The uncertainty of the determined edge locations is about one pixel. Finally, we performed a linear fitting of the red diamonds. The average upflow speed was estimated from the slope of the fitted line, which turns out to be 19.08$\pm$1.89 km s$^{-1}$. To obtain the period of the oscillating upflow, we first produced S-T diagrams for several cuts across the upflow at different heights. One cut is shown as the red horizontal line in Figure 2A, and the corresponding S-T diagram is presented in Figure 2D-F. We then determined the central position of the oscillating feature by applying a Gaussian fitting of the intensity profile across the upflow at each time step. The measurement error of AIA  intensity is dominated by the Poisson noise \citep{2012SoPh..275...41B},  which can be approximated by the square root of the photon number at each pixel. Considering this measurement error, the 1$\sigma$ fitting error of the determined central position at each time step can be calculated. The central position clearly reveals a periodic oscillating behavior, and a sinusoidal fitting yielded an oscillation period of 12.87$\pm$0.33 min. Note that the uncertainties of the flow speed and period determined here represent the 1$\sigma$ fitting errors. 

A close inspection of the image sequences suggests that this upflow is related to the motion of a SAD. Figure 3A and B show that the upward wavelike or oscillating plasma motion (in 94~\AA~images) is preceded by the arrival of a SAD (in 131~\AA~images) at 11:18 UT. To determine the speed of the SAD, we stacked several images with a constant time interval and tracked the downward motion of the faint SAD manually. The red circles in Figure 3B mark the tip of the downward moving SAD. The diameter of the red circles (8 pixels) was taken as the measurement error of the SAD location. By performing a linear fitting of the determined SAD locations, the speed of the SAD was estimated to be 24.54$\pm$5.84 km/s. We also measured the width of the SAD at a location just above the upflow. In Figure 3E, we show the 131~\AA~intensity across the SAD (along the cut S1 in Figure 3D). Again the error of the intensity is essentially the Poisson error. A Gaussian fitting of this intensity profile yielded a diameter of 6.90$\pm$0.80 Mm (full width at half maximum) for the SAD.

\begin{figure*}
\label{fig3}
\begin{overpic}[width=\textwidth]{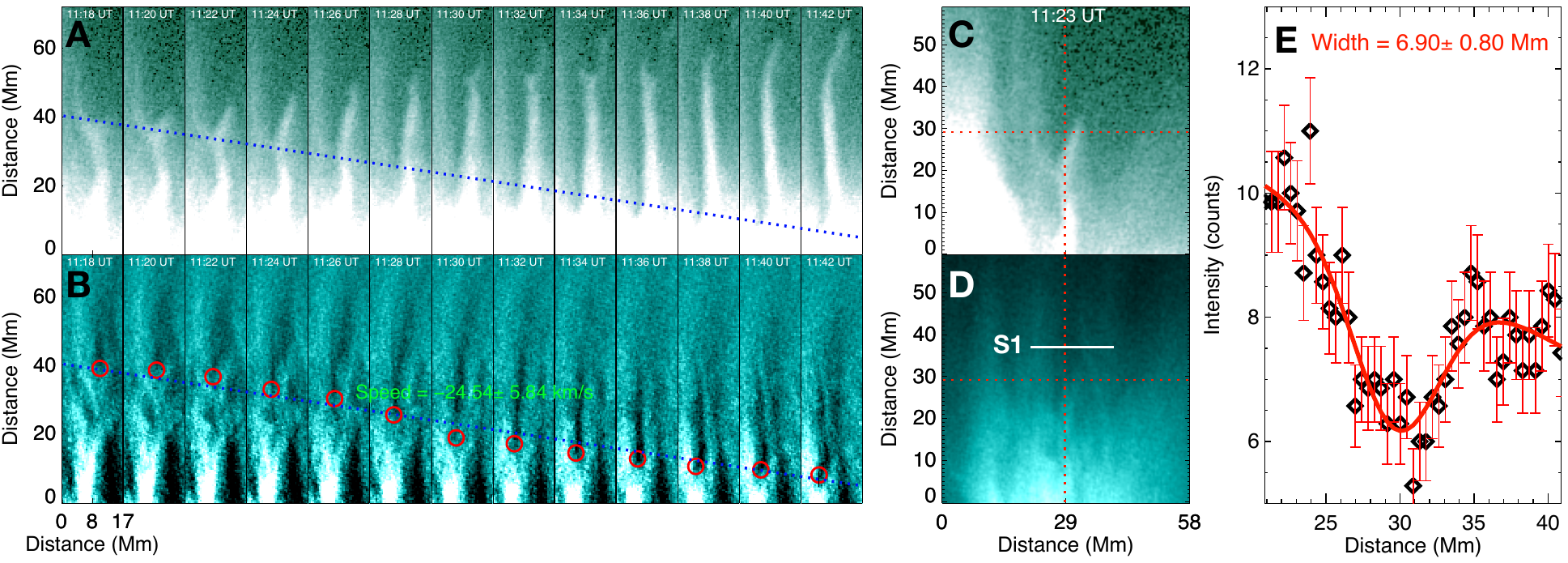}
\end{overpic}
\caption{Association of the wavelike upflow with a SAD. A: temporal evolution of the upflow in AIA 94~\AA. B: temporal evolution of the associated SAD in AIA 131~\AA. The AIA 131~\AA~ images have been enhanced by subtracting a smoothed (over 50-pixels along the X-axis) background at each Y-pixel. The red circles mark the tip of the downward moving SAD. The blue dashed line represents a linear fit of the locations of these red circles. The same line is also overplotted in the AIA 94~\AA~images in panel A. C-D: AIA 94~\AA~image of the upflow and 131~\AA~image of the SAD at 11:23 UT. The two red dashed lines (vertical and horizontal) are overplotted to demonstrate the coincidence of the upflow and the SAD. E:  AIA 131~\AA~intensity (diamond) and the associated measurement error (red vertical bar) along the cut S1 shown in panel D. The red line shows the Gaussian fit (with a linear background). }
\end{figure*}
%\begin{figure}
%\label{fig4}
%\begin{overpic}[width=.47\textwidth]{Fig4_re.pdf}
%\end{overpic}
%\caption{ A cartoon showing the possible origin of the observed vortex shedding in the post-flare supra-arcade region. It is a combination of two figures from Savage \& McKenzie 2011 \cite{2011ApJ...730...98S} (bottom part) and Homann 1936 \cite{Homann1936} (top part).}
%\end{figure}

The Strouhal number ($St = L/PV$)  of this propagating upflow can be computed using the three parameters: size of the blunt body ($L$), period of the oscillation ($P$) and speed of the flow ($V$). In our case the shrinking post-flare loop possibly acts as the blunt object, and the SAD is likely the wake of the shrinking loop  \citep{2012ApJ...747L..40S}. Considering this, the blunt body size can be reasonably approximated by the width of the SAD. Since the blunt body is moving against an upflow, the speed $V$ should be the relative speed between the upflow and the downward moving SAD. Taking the measured values of these parameters, we obtained a Strouhal number of 0.20 $\pm$ 0.06. It is worth noting that here we used the flow velocity projected onto the plane of the sky, which is usually smaller than the full velocity. To evaluate the impact of this effect on the calculated Strouhal number, we examined the coordinate of the active region several days before the flare, when the active region was located on the solar disk. By assuming that the location of the active region did not change too much as it rotated to the limb, we found that the angle between the radial direction at this active region and the line of sight was about 90 deg in our observation. So there is essentially no line-of-sight projection effect if we reasonably assume that the upflow and SAD both move in the radial direction. The projection effect is negligible even if there is a small deviation from the radial direction, since a 10-deg deviation can lead to only $\sim$1.5\% difference of the flow velocity.

Oscillations observed in the solar corona are generally interpreted as signatures of different MHD waves \citep{2004psci.book.....A,2005LRSP....2....3N, 2004ASSL..317..283M}. Slow magnetoacoustic waves often propagate with the speed of sound at a particular temperature. Other MHD waves such as fast kink waves, fast sausage waves and Alfv$\acute{\text{e}}$n waves propagate at a speed of the order of the Alfv$\acute{\text{e}}$n speed. The sound speed, calculated as $C_{s} = 0.152 \times T^{1/2}$ km s$^{-1}$, is around 370 km s$^{-1}$ at a temperature of 6 MK. The phase speeds of kink waves reported in coronal current sheets and postflare supra-arcade loops are in the range of 100--700 km s$^{-1}$ \citep{2005A&A...430L..65V,2016ApJ...829L..33L}. The propagation speed of our observed oscillating feature is $\sim$19 km s$^{-1}$, which is 1--2 orders of magnitude lower than the typical sound speed and Alfv$\acute{\text{e}}$n speed in the corona. This difference indicates that the wavelike upflow has a different origin. 

A wavelike pattern could also originate from the shearing motion between two fluids due to Kelvin-Helmholtz instability (KHI). Some of the wavelike patterns observed in magnetized fluids, e.g., solar atmosphere \citep{2010ApJ...716.1288B, 2010SoPh..267...75R, 2013ApJ...766L..12M,2013ApJ...774..141F, 2016ApJ...830..133K, 2018NatSR...8.8136L,2018SoPh..293....6F}, comet tails \citep{1980SSRv...25....3E}, planetary atmospheres \citep{2004Natur.430..755H,2010JGRA..115.7225M,2010Sci...329..665S}, and astrophysical jets \citep{1984RvMP...56..255B,2001Sci...294..128L} are interpreted as being caused by the KHI. However, KHI occurs when the velocity difference of two fluids moving in parallel exceeds a critical value, which is of the order of twice the Alfv$\acute{\text{e}}$n speed if the flows are field aligned \citep{1980PhFl...23..939L}. Since we do not know the angles between the magnetic fields and fluid velocities, we cannot determine the critical value in our case. Thus, it is unclear whether the KHI can explain the observed oscillation or not.

On the other hand, a Strouhal number of the order of 0.2 is a robust indicator for vortex shedding in rarified plasmas \citep{2010PhRvL.105e5004G}. The Strouhal number calculated from the measured physical parameters in our observation is 0.20 $\pm$ 0.06. This Strouhal number is consistent with the prediction from a numerical simulation of Alfv$\acute{\text{e}}$nic vortex shedding \citep{2010PhRvL.105e5004G}, which is generally in the range of 0.15--0.25. Note that this simulation considered the interaction of an initially uniform and steady plasma flow with a cylindrical blunt body in the adiabatic case. The possible slight difference between our measured value and the prediction may be related to the deviation from the ideal situation of the simulation. Our observation thus supports the presence of vortex shedding in the solar corona. 

What causes the oscillation in the flow in the reconnection current sheet? Without sufficient resolution and signal-to-noise ratio of the observation, this question cannot be easily answered. However, the AIA observation shows a close relationship between the downward motion of a SAD and the initiation of the upward wavelike motion. Based on our understanding of SADs, we propose the following possible scenario to explain the generation of vortex shedding.Because of the magnetic tension, a newly formed postflare loop shrinks and acts as a blunt object moving downwards through the supra-arcade region. The wake of the shrinking loop exhibits as a SAD \citep{2012ApJ...747L..40S}. The  cross section of the loop generates periodic vortices due to the pressure difference in the wake of the shrinking loop and the surrounding plasma, leading to a wavelike plasma upflow behind the shrinking loop. This scenario might be similar to a wavelike flag in the wind. The fact that the oscillating upflow starts from the head of a SAD in our observation is consistent with this scenario. Though the presence of vortices cannot be unambiguously identified from the diffuse and weak emission of AIA 94~\AA, vortex motions have indeed been previously reported in similar postflare supra-arcade regions \citep{2013ApJ...766...39M,2016ApJ...831...94S}. Additionally, the properties of oscillating flows caused by vortex shedding depend on the Reynolds number, and vortices might not be clearly observed in some cases \citep{Homann1936}. We also noticed that nonlinear development of the KHI may lead to vortex shedding in fluids \citep{bookRodi2005}. Future studies are needed to investigate the detailed formation process of vortex shedding in the magnetized and turbulent supra-arcade regions \citep{2013ApJ...766...39M}.

Nevertheless, our work suggests that vortex shedding could exist in the magnetized solar atmosphere. Vortex shedding has been widely recognized to be an important physical process. For example, strong flows or winds could more easily destabilize ships or tall chimneys if vortex shedding is involved. Studying the effect of vortex shedding in MHD conditions, i.e., in laboratory plasmas \citep{Aydemir2005,Dousset2008}, in the Sun \citep{Emonet_2001}, or some other astrophysical conditions \citep{Ott1978,Jones_1999}, could also shed new light into some unresolved physical problems. For instance, vortex shedding might explain the origin of some oscillations observed in the solar atmosphere \citep{2009A&A...502..661N}. Such oscillations may lead to destabilization of coronal structures,  which may eventually result in solar eruptions. They may also play an important role in the process of energy transport in the solar atmosphere.

\begin{acknowledgments}
This work is supported by NSFC Grants No. 11825301, 11790304(11790300), and No. 11850410435, the Strategic Priority Research Program of CAS (Grant No. XDA17040507), the Strategic Pioneer Program on Space Science of CAS (Grants No. XDA15011000 and No. XDA15010900), and the Russian Foundation for Basic Research (Grant No. 18-29-21016). H.T. acknowledges support by ISSI/ISSI-BJ to the teams ``Diagnosing heating mechanisms in solar flares through spectroscopic observations of flare ribbons" and ``Pulsations in solar flares: matching observations and models." We thank Jiansen He, Tom Van Doorsselaere, and Vaibhav Pant for helpful comments and suggestions. 
\end{acknowledgments}


%These movies are available at DOI:https://doi.org/10.1103/PhysRevLett.123.035102
%\bibliography{myreferences}

\end{document}